# ANOMALOUS GRAVITATIONAL EFFECTS IN THE UNIVERSE

## JOHANN ALBERS

Fachbereich Physik der Universität des Saarlandes
66041 Saarbrücken, Germany
e-mail: albers@rz.uni-sb.de

## ABSTRACT

Based on previously published alternative reflections on gravitation, additional conclusions concerning anomalous gravitational effects in the universe are derived. For systems with spherical mass distribution and high central density of the luminous matter, abnormal increases in the gravitational potential towards the centers of these systems are derived. These anomalous increases are calculated on the basis of the amount and distribution of the observed luminous matter. For spiral galaxies, the calculation yields the astonishing result that the anomalous part of the gravitational potential is not produced by the large mass of the spiral system but by the small mass of the central bulge. A noticeable anomalous increase can only be expected close to the center, e.g. in the innermost region with a diameter on the order of only 1 pc in a bulge with a radius of 500 pc. Experimental observations are cited reporting such anomalous gravitational effects in the four spherical systems considered: globular star clusters, the bulges of spiral galaxies, elliptical galaxies, and clusters of galaxies. Conventionally, on the basis of Newtonian Mechanics and the amount of luminous matter, these effects are unexpected. They are most commonly explained by assuming the existence of appropriate large black holes or a sufficient amount of dark matter.

Subject headings: alternative theory - gravitation - black hole - dark matter - globular clusters - elliptical galaxies

## 1. INTRODUCTION

In this paper, some special points included in a foregoing article about alternative reflections on gravitation will be discussed in more detail. For this purpose, the main points of the alternative reflections (Albers 1997a) are recapitulated briefly to the extent necessary to understand the focus of this paper: Every mass emits and absorbs gravitational radiation which is treated in complete analogy to the laws of optics and electromagnetic radiation. If the effect of all masses $m_j$ of the universe is summed up for a reference point Q, one obtains a certain intensity $I_0$ of the gravitational radiation

$$I_0 = \Sigma_j(Em_j/R_j^2). \qquad (1)$$

Here, $R_j$ is the distance between the location of $m_j$ and the reference point Q, and E is a constant that describes the efficiency to emit gravitational radiation per unit mass.

Two masses $m_1$ and $m_2$ near Q are considered which both partially absorb $I_0$. The absorption leads to a momentum transfer and thus to forces directed towards the centers of both masses. By its own absorption, every mass shields the other one partially against the radiation $I_0$, leading to an attractive force between both masses, which follows exactly Newton´s law of gravitation

$$F = -G_f m_1 m_2 / r^2. \qquad (2)$$

However, $G_f$ is not the „universal constant of gravitation", G, but a factor of proportionality, denoted as gravitational factor in the following, which is linearly dependent on $I_0$

$$G_f \propto I_0. \qquad (3)$$

From equations (1) and (3), the following relation can be deduced

$$G_f \propto \Sigma_j(m_j/r_j^2). \qquad (4)$$

Thus, the attractive force in equation (2) is a secondary effect which is based on the existence of all masses $m_j$ in the universe.

The primary interaction between $m_1$ and $m_2$, however, is a force of repulsive nature. It is due to the momentum transfer produced by the absorption of gravitational radiation emitted by each of the two masses and absorbed by the other one. A suitable system to explain this effect is a system with spherical mass distribution, a mass M and radius R. According to equation (1), its masses produce gravitational radiation with a directional component $I_d$ at its periphery

$$I_d \propto M/R^2. \qquad (5)$$

It has previously been shown (Albers 1997b) that the balance between primary, repulsive forces due to $I_d$ and secondary, attractive forces due to $I_0$ can stabilize highly concentrated spherical mass accumulations. It has been indicated that such objects can be observed in the universe in the form of globular star clusters, elliptical galaxies and spherical clusters of galaxies. For these objects, the expected linear dependence of their mass M on the square of their diameter R could be revealed

$$M/R^2 = \text{constant}, \qquad (6)$$

and the value was determined to be about 160 $M_{Sun}/pc^2$. The above derivation is a short summary of the points in (Albers 1997a, 1997b) relevant to the conclusions in this study. The following is based mainly on equation (4).

## 2. Spherical Mass Accumulations

In the introduction, three spherical systems treated previously (Albers 1997b) were mentioned. A member of a fourth system of spherical mass accumulation should be added since exceptionally reliable data has been published concerning the gravitational potential near its center: the bulge of our Galaxy. The bulges of spiral galaxies were not taken into account earlier, although also these systems seem to be old and stable, with old stars of population II. It is, however, difficult to get reliable values of M and R for the bulges alone, because the luminosity and apparent diameter cannot easily be attributed to either the bulges or the spiral arms. The following data are taken from a textbook (Scheffler & Elsässer 1992, p. 429 ): from R = 500 pc and a mean surface mass density of 100 $M_{Sun}/pc^2$ , a value of 314 $M_{Sun}/pc^2$ can be calculated for $M/R^2$ in equation (6). If added to the data of the other three systems (Fig. 4 in Albers 1997b), it becomes apparent (Fig. 1) that this member of the fourth spherical system is very compatible with the previous data. The deviation from the straight line which corresponds to $M/R^2 = 160$ $M_{Sun}/pc^2$ is below a factor of 2, thus comparable to the deviations from the straight line in the other systems.

On the right side in Fig. 1, three points are added which correspond to the $M/R^2$ - value of the universe. The upper point results from a radius of 4 Gpc and a mass according to the commonly cited $10^{11}$ galaxies with $10^{11}$ suns with a mean mass equal to that of our sun. The slightly lower value corresponds to the data cited earlier (Albers 1997a): R = 4 Gpc and a homogeneous density of $5\times10^{-28}$ kg/m$^3$. The lowest point corresponds to a value of $2\times10^{-28}$ kg/m$^3$ for the baryonic matter in the universe (Bergmann-Schäfer 1997, p. 441). These three points are not necessarily expected to lie on the straight line since they represent $I_0$ in equation (1) which is responsible for the secondary, attractive interaction, whereas all other data points represent the primary, repulsive interaction ($I_d$ in equation (5)). They do, however, limit the region where further spherical systems may be expected beyond the clusters of galaxies. In such systems, e. g. superclusters of galaxies, 10 to 100 times larger in diameter than clusters of galaxies (Bergmann-Schäfer 1997, p. 299), the perfect spherical symmetry can hardly be expected. Their diameter R is very large and their mean spatial density decreases proportional to 1/R (equation (6)). Furthermore, their relaxation times are at least on the order of the life time of the universe, so that their recovery after a disturbance by the gravitational influence of other masses takes long enough to prevent the perfection of spherical symmetry and radial profile of these large systems.

A very important point for the following calculations is the radius-dependent spatial mass density inside the spherical systems. As discussed previously (Albers1997a), the density dependence on the distance from the center of the spherical system, r, must be stronger than 1/r:

$$\rho \propto r^{-a}, a > 1 \qquad (7)$$

An exponent a = 1 already follows from the assumption that not only at the periphery but also inside every sphere with radius r < R, the mass m(r) included in the sphere with radius r creates the necessary primary repulsive force against the incorporation of additional masses into the sphere with radius r, according to $I_d$ in equation (5)

$$I_d = M(r)/r^2. \qquad (8)$$

Reasons discussed earlier (Albers 1997a) strengthen the increase of $\rho$ towards the center. However, the exact determination of the exponent a due to these effects, and perhaps the derivation of a new function more suitable than equation (7), requires a more quantitative description of all processes. Furthermore, every real mass accumulation will show more or less strong deviations from the exact functional description of the spherical state of equilibrium. Therefore, the following calculations will be based on the power law in equation (7) with an exponent a = 1.

These considerations about the radius dependent density inside spherical objects may be compared with experimental observations and their published mathematical descriptions: Recent observations of M 15 with high angular resolution by the Hubble Space Telescope (HST) (Guhathakurta et al. 1996, Fig. 13) allow the determination of the spatial mass density profile near the center of this globular cluster. In the innermost region, from 0.02 pc up to about 0.08 pc, the mass density can be approximated by a power law with an exponent a =

1.14, compatible with the above considerations. The E5-galaxy NGC 3377 should be mentioned as an additional example, for which the deprojected brightness profile is shown graphically down to log r (arcsec) = -1.8 by Kormendy et al.(1998, Fig. 6). A fit to the innermost part of this curve with equation (7) yields a value of a = 1.3.

Tremaine et al. (1994) describe a one-parameter family of models for spherical stellar systems with power-law density cusps, as given by equation (7), at the center. The exponent a in their models extends from 0 to 3 thus involving two of the most successful analytic models for elliptical galaxies and bulges of disk galaxies: a = 2 (Jaffe 1983) and a = 1 (Hernquist 1990). These examples of fits and analytical models may be sufficient to show that the conclusions developed on the basis of the alternative reflections on gravitation about the radius-dependent density are compatible with experimental observations and their empirical functional description, and that the value 1 of the exponent a is reasonable within the broad field of reported parameter values.

### 3. EXPECTED ANOMALOUS GRAVITIONAL EFFECTS

There are fundamental differences between the conventional treatment of gravitational effects based on Newtonian mechanics (standard gravity theory in the following) with a constant value of the „universal constant of gravitation", G, on the one hand, and the gravitational factor $G_f$ derived from the alternative reflections on gravitation on the other hand. The value of $G_f$, calculated according to equation (4), may be tracked along the path of the reference point Q, which is moved through the universe. Inside huge voids, $G_f$ will accept low values, whereas it takes high values in regions where clusters of galaxies with their huge mass accumulations are concentrated.

A reference value of $G_f$ according to $\Sigma_j(m_j/r_j^2)$ in equation (4), expressed in units of $(M_{Sun}/pc^2)$, may be derived from the mass M and the radius R of the universe. These values, however, spread over a wide range depending on the data source. Extremely high estimated masses M of $10^{23}$ to $10^{25}$ $M_{Sun}$ (Bergmann-Schäfer 1997) are usually based on the assumption that the luminous matter makes up only a few percent of the total mass of the universe, whereas the larger part is expected to exist in the form of dark matter. In this study, the above mentioned conservative value of M = $10^{22}$ $M_{Sun}$, based on the luminous matter of $10^{11}$ galaxies each containing $10^{11}$ suns, will serve as a basis. With R = 4 Gpc as the radius of the universe, it follows that $M/R^2$ = 625 $M_{Sun}/pc^2$, a value already documented in Fig. 1. Following the common assumption of a homogeneous distribution of this mass inside R, equation (4) leads to a contribution $G_u$ = 1875 $M_{Sun}/pc^2$ (equal to $3M/R^2$) of the universal masses to the gravitational factor $G_f$. If, however, the reference point Q is located at a point of increased mass density, for example inside a spiral galaxy, additional contributions $G_s$ and $G_b$, resulting from the masses in the spiral system and the masses in the bulge, respectively, raise the value of $G_f$ above this mean value $G_u$:

$$G_f = G_u + G_s + G_b. \qquad (9)$$

In order to examine the properties of the three different terms in equation (9), the own Galaxy may serve as an example. For the bulk, the above mentioned values R = 500 pc and a mean surface mass density of 100 $M_{Sun}/pc^2$ are used. As discussed earlier, a 1/r dependence of the mass density is assumed inside R. For simplicity, it is further assumed that the density outside R is zero and that the overall mass of the bulge ($7.85 \times 10^7$ $M_{Sun}$) is located inside R. If, for the calculation of $G_f$, the summation of the term on the right side of equation (4) is performed for all masses $m_j$ lying between R and r with r < R, it follows that the sum diverges proportional to ln (R/r) for the point of reference at r=0. This means that, at the center of the bulge of our Galaxy as well as at the center of all other mentioned spherical objects, the gravitational potential and the interactions according to Newton´s law of gravitation in equation (2) are determined by an infinitely high gravitational factor $G_f$ , instead of the „universal constant of gravitation“, G, with the same value at every place. Of course, this result is only of theoretical interest due to the limited values of real density and of observational resolution near the centers of these objects. However, further conclusions may be drawn which are closer to experimental results. The bulge component $G_b$ of the gravitational factor $G_f$ according to equation (9) is calculated numerically as described by equation (4) for points Q inside R and at a distance r away from the center of the bulge of our Galaxy. The results are shown in Fig. 2, and the value $G_u$ = 1875 $M_{Sun}/pc^2$ is indicated as a horizontal line. Additionally, a third curve is drawn that describes the radiant dependent third contribution $G_s$ of the spiral system. For simplicity, the spiral system is approximated by a plate with a thickness of 1000 pc, limited towards the center of the Galaxy by the radius $R_b$ = 500 pc of the bulge and bound on the outside by the commonly reported radius of the Galaxy of about $R_s$ = 15 kpc. For the mass which is assumed to be distributed homogeneously within the spiral system, values can be chosen between the frequently mentioned $10^{11}$ $M_{Sun}$ and the value resulting from the luminosity of the Galaxy of $1.4 \cdot 10^{10}$ $M_{Sun}$ (Bergmann-Schäfer 1997). The following will employ the latter value since its derivation is not based on the viral theorem from observations of gravitational effects. The spiral component $G_s$ of the gravitational factor $G_f$ has about the same magnitude as the contribution $G_u$ of the universe and shows no strong variation inside the radius R = 500 pc of the bulge, as can be seen from Fig. 2. The bulge contribution $G_b$, however, shows a dramatic increase towards the center, as can be seen from the data inside the innermost 25 pc in Fig. 3. At about 2.5 pc, $G_b$ reaches a value equal to the sum of $G_s$ and $G_u$ and increases more steeply with decreasing distance from the center. This indicates an anomalous gravitational effect, characterized by a diverging increase towards the center of the luminous matter. This anomalous effect is produced by the small mass of the bulge and not by the mass of the spiral which is more than two orders of magnitude higher.

Of course, such dramatic anomalous gravitational effects are a priori not expected from standard gravity theory, although the standard theory can explain such behavior in the presence of a black hole in the center. But there are at least two points where the interpretations of observed anomalous gravitational effects should differ. On the basis of the alternative reflections, no typical X-ray emission and no Keplerian decrease of the rotational velocity are expected near the center, two effects which are characteristic for the existence of a black hole.

In Fig. 4, the results of further numerical calculations are shown which were performed in the same way as described in connection with Fig. 2, but for spherical objects with radius R between 1 and $10^6$ pc, which covers the range of the four different groups of spherical objects in Fig. 1. The gravitational factor $G_f$ at the ordinate consist of only two components, the contribution of the universe, $G_u$, again taken as a constant with a value of 1875 $M_{Sun}/pc^2$, and the value $G_{sp}$ of the spherical object. The spherical objects are defined by their radius R and their mass M, following from the value $M/R^2$ = 160 $M_{Sun}/pc^2$ which represents the straight line in Fig.1. At distances r comparable to R, the contribution $G_{sp}$ of the spherical object begins to increase with decreasing r. Inside r = R/2, this increase takes place in a similar manner for all objects, linearly with the logarithm of the distance r from the center. At about r

= R/200, the contribution of the spherical object reaches that of the universe. At smaller distances, the gravitational effects are mainly determined by the strong, anomalous increase of $G_{sp}$, produced by the masses of the spherical object. Thus, in analogy to the expectations discussed above in connection with the bulge of our Galaxy, similar anomalous gravitational effects are expected for all four groups of spherical objects shown in Fig. 1 due to the amount and distribution of the luminous matter. With the standard gravity theory, these effects can only be explained by assuming the existence of unseen matter which contributes to the gravitational potential near the center.

## 3. OBSERVATIONAL TECHNIQUES

The best known method to determine gravitational effects is the measurement of rotational velocities in the outer region of spiral galaxies. The rotational velocity v at a distance r from the center of a radial symmetrical mass accumulation is related to the mass inside the sphere with radius r by

$$v^2 = GM(r)/r. \qquad (10)$$

If the entire mass M is concentrated within r, a condition which is usually fulfilled when observing black holes, a Keplerian decrease proportional to $r^{-0.5}$ of the rotational velocity v is expected. This method was used in more than one thousand experiments and lead to the conclusion that every spiral galaxy contains dark matter, especially in the region outside the galactic bulge. The four systems discussed here normally show no pronounced rotational but mainly vibrational movements of their components. Therefore, the determination of the velocity dispersion is the appropriate method to get information about the gravitational potential inside these systems. In galaxy clusters with spherical symmetry, the mass may also be determined on the basis of X-ray measurements since this method typically yields errors of less than about 15% (Schindler 1998). However, for the considerations in this study, the velocity dispersion data remain the best source of information. The velocity dispersion at the center of a spherical system with mass M and radius R is proportional to the square-root of (MG/R). Assuming the relation $M/R^2$ = constant according to the straight line in Fig. 1, it follows that the velocity dispersion is proportional to the radius R of the spherical object. Due to the experimental limitations there may be assumed a constant value of the limit of error in the determination of the velocity dispersion. Then it follows that the relative error in the determination of the gravitational potential decreases with increasing diameter from the globular clusters towards the clusters of galaxies. In order to evaluate the gravitational potential near the center of our Galaxy, the proper motions with their components along all three spatial directions of single stars were determined during a period of several years (Eckart & Genzel 1996). The conclusions resulting from these experimental methods are mentioned in the following chapter.

## 4. OBSERVED ANOMALOUS GRAVITATIONAL EFFECTS

For each of the four previously mentioned spherical mass accumulations, at least one example with anomalous gravitational potential will be mentioned in the following. The cited references contain not only supporting measurements, e.g. angular resolved velocity dispersion data, but also explicit indications that anomalous gravitational effects exist. These may be explained conventionally by an increase of the M/L ratio or by the presence of MDOs (massive dark objects), i.e. black holes or dark matter.

**Globular clusters**: These are the smallest spherical systems of mass accumulations discussed in this paper. Due to their low masses, these systems possess low values of the velocity dispersion with high relative limits of error. As observed by Gebhardt et al. (1997) in M15, the velocity dispersion profile with values of about 11 km/s „can be equally well represented either by a stellar population whose M/L varies with radius from 1.7 in solar units at large radii to 3 in the central region, or by a population with a constant M/L of 1.7 and a central black hole of 1000 $M_{Sun}$". These results confirm the expected behavior according to chapter 3. Due to their observations of the velocity data of the same object, Dull et al.(1997) conclude that „there appears to be no need to invoke the presence of a massive central black hole in M15". In their opinion, the anomalous gravitational effects may just as well be explained by the mass of about $10^4$ neutron stars. The existence of these dark objects, however, is probably as difficult to prove as that of the above mentioned black hole.

**Bulges of spiral galaxies:** In this system, the previously mentioned observations of Eckart & Genzel (1996) may be the most important. They assume that the enormous increase of the proper motions of stars near the center of our Galaxy must be ascribed to Keplerian orbitals around a black hole, the mass of which they calculate as $2.45 \times 10^6$ $M_{Sun}$. Already much earlier (Townes & Genzel 1990), it was clear from the motions of stars and clouds that the anomalously high velocities lead to discrepancies between theory and observation only within a few light-years around the center. This small distance is rather well compatible with the data and conclusions in connection with Fig. 3.

Ghez et al. (1998) measured the proper motions near the Galactic center with even higher angular resolution, and they also derive a central dark mass with a similar value of $2.6 \times 10^6$ $M_{Sun}$. The uncertainties in the measurements mathematically allow for the matter to be distributed over a volume with a diameter of at most $10^{-2}$ pc. But because no realistic cluster seems to be physically tenable, these authors, too, come to the conclusion that our Galaxy harbors a massive black hole.

In our nearest neighbor galaxy, the spiral galaxy M31 which has similar properties as our own Galaxy, the observed anomalous gravitational effects were explained by a mass-luminosity ratio M/L which increases strongly when approaching the center. Of course, one can calculate M alone only if one assumes a constant G, otherwise the product GM can not be separated into its factors. If one takes $G_f M/L$ instead of GM/L, it becomes clear that the conventionally determined increase of M/L must occur near the center. In M31, this happens inside a diameter of about 3 pc (fig.5 in Kormendy & Richstone 1995), a value rather well compatible with the expectations from Fig. 3.

Magorrian et al. (1997) present a demography of massive dark objects in the centers of 36 nearby galaxies, including spiral galaxies. They conclude that a fraction of 97% contain massive dark objects where the masses $M_D$ of these MDOs seem to be correlated with the masses $M_{Bulge}$ of the bulge of these galaxies. Typically, the masses of the MDOs are 0.006 times that of the bulges. The masses $M_{Bulge}$ themselves may be considerably smaller than the total mass of the galaxy. This fact and a comparison with elliptical galaxies have already been described by Rees in 1990 (Rees 1990): „The mass that accumulates in the center of an elliptical galaxy may be proportional to the total mass of the galaxy. For spiral „disk" galaxies like the Milky Way, the hole may be related not to the total galaxy mass but to the mass of the dense central bulge of stars which is much smaller than a typical elliptical galaxy". This is just the result that is expected on the basis of the estimations performed in chapter 3.

The assumed black holes at the centers are not the sole candidates of unseen masses in spiral galaxies. Due to the easily measurable anomalous gravitational effects outside the bulges, the spiral galaxies demand a second type of unseen mass with different radial symmetry: dark matter with a distribution appropriate to explain the non-Keplerian rotation velocities in the spiral systems.

**Elliptical galaxies**: Black holes are found at the centers of small as well as of huge elliptical galaxies. Within the dwarf elliptical galaxy NGC 4486B (with an absolute magnitude M of only -16.8 mag), the velocity dispersion increases from 116±6 km/s in the outer regions to 281±11 km/s at the center (Kormendy et al. 1997). Compared to the data of the small systems of globular clusters, the velocity dispersion value of NGC 4486B is far above its limit of error. The measured data can be fitted best by a black hole with a mass of about $6 \times 10^8$ $M_{Sun}$.

The giant elliptical galaxy M87, about one hundred times brighter than NGC 4486B, presents velocity dispersion data which are consistent with the existence of a $3.2 \times 10^9$ $M_{Sun}$ black hole at its center (Macchetto et al. 1997). According to Reynolds et al. (1997), it is believed that most giant elliptical galaxies possess nuclear black holes with masses in excess of $10^8$ $M_{Sun}$. It might then be expected that quasar-like luminosities even from the nuclei of quiescent elliptical galaxies are produced. In their opinion, it is a puzzle that such luminosities are not observed.

Recent examinations by van der Marel (1998) are concerned with a collection of 46 early type galaxies (E, E/S0 or SO) for which highly angular resolved surface brightness profiles are available from HST-observations. The measured profiles are compared with model profiles proposed earlier by Young (1980). In the innermost region, between 0.1 arcsec and about 1 arcsec away from the center of the luminous matter, differences between these profiles usually occur which are explained by the existence of black holes with appropriate masses, located at the center of the luminous matter. In those cases where a comparison is possible with kinematically determined black hole masses, good agreement is usually reached between both values. In the opinion of the author, the results support the hypothesis that every galaxy (spheroid) has a central black hole. The most interesting point in connection with these results, however, is the fact that the anomalous gravitational effects, even if they are explained conventionally by the assumption of black holes, can be calculated on the basis of amount and distribution of the luminous matter in the spherical or nearly spherical objects. Just this is expected when, following the derivation in this paper, the anomalous gravitational effects are described by the gravitational factor $G_f$ which is, according to equation (4), dependent on the amount and distribution of the luminous matter around the point under consideration.

**Clusters of galaxies:** The spherical systems discussed until now, are normally components of still larger systems, i.e. the clusters of galaxies. Therefore the data determined from the smaller systems could be used to describe the gravitational relations inside the clusters of galaxies. These systems are so large that the observation of their anomalous gravitational effects has been possible for a long time. It seems to be clear that the gravitational potential shows an anomalous increase towards the center of many (Sciama et al. 1992, Sciama 1993), if not of all clusters of galaxies. In these cases, however, mainly due to the huge diameter, the effect is not attributed to the existence of a black hole but to the presence of huge amounts of dark matter with the appropriate spatial distribution. Clusters of galaxies possess long relaxation times, and a coarse grained structure due to the small number of components, about $10^3$ to $10^4$ galaxies. The three smaller spherical systems, globular star clusters, bulges of spiral galaxies and elliptical galaxies, however, contain between $10^5$ and $10^{22}$ stars as components. Thus it seems clear that in clusters of galaxies, which often contain a huge elliptical galaxy at their center, the luminous matter is not distributed as regularly as in the smaller spherical systems. This can depress the increase of the gravitational factor $G_f$ towards the center and favour the dark matter concept instead of a huge black hole in the conventional description.

## 5. DARK MATTER AND BLACK HOLES

The alternative reflections on gravitation are obviously suitable to indicate places in the universe where anomalous gravitational effects occur which, together with the amount of the observed luminous matter, cannot be explained by Newtonian mechanics. However, this does not preclude that the usually accepted sources of increased gravitational effects, black holes and dark matter, also contribute to the observed anomalous gravitational effects in these places. Although this is not the place for a broad discussion of the enormous number of publications on the subject, a few recent papers should be mentioned that strongly speak against the existence of these gravitational sources at several places.

According to Mateo (1993), the dark matter problem has existed for more than 60 years, since Zwicky first demonstrated that galaxy clusters contain significant amounts of dark matter. Despite the huge efforts to prove the existence of dark matter by properties other than the observed anomalous gravitational effects, especially during the last decade, no clear evidence of its existence has been established. The entirely different linking of dark matter to the masses in spiral galaxies on the one hand, and to the masses of clusters of galaxies on the other hand, as has been documented by Sciama et al. (1992), does not contribute to the acceptance of this concept.

However, observations seem to exist that directly contradict the existence of dark matter at places where it should be present in large amounts. MACHOs (massive astrophysical compact halo objects) are one of the mainly assumed constituents of dark matter. Observations by the Hubble Space Telescope of regions inside three rich Abell clusters of galaxies where the amount of dark matter should exceed the luminous matter considerably - often, factors of more than 100 are reported - , lead to the conclusion that the contribution of MACHOs lies below a negligible amount of about three percent of the luminous matter (Boughn & Uson, 1997). Hellemans (1997) summarizes efforts to detect unseen masses in the own galaxy in a notice with the title: „Galactic Disk Contains no Dark Matter". Other tests for probing dark matter in the own galaxy in the form of the expected but not found amount of MACHOs lead Burrows and Liebert (1995) to the conclusion: „All too often, discussions of the halo dark matter have resembled mediaeval discourses on the Aristotelean quintessence or the angelic population of the empyrean. Astronomers seemed to be involved in shadow boxing with a Nature jealous of its secrets". As a last contribution concerning MACHOs, recent statements by Glanz (1998a) may be cited here: „ The dark objects thought to inhabit the Milky Way's halo, accounting for its missing mass, may actually be dim stars in nearby galaxies".

Dust, subatomic particles, neutrinos, axioms et cetera belong to the second group of possible candidates for dark matter at the lower end of the mass scale. The nearest and therefore perhaps best place for experiments to estimate their possible contribution to the dark matter masses is our solar system. At a distance of about 8.5 kpc from the Galactic center, the non-Keplerian rotation curve demands an appreciable amount of dark matter. Besides the known „dark matter" of the planets, one gets a negligible amount of less than one earth mass for the dark matter around the sun interior to Neptune's orbit (Anderson et al. 1995). This results in an extremely low bound of only $3 \times 10^{-6}$ of the dark matter compared to the luminous matter of the sun. But despite this and many other papers reporting the absence of different types of dark matter at different places in the universe, the existence of dark matter as the cause of anomalous gravitational effects has not been ruled out with certainty yet.

The second conventionally adopted source to explain the anomalous gravitational effects, the black hole, of course possesses the necessary radial symmetry. But the expected typical X-ray emission which is characteristic of Cygnus X-1, the most prominent candidate for a black hole, is missing in our Galaxy (Grindlay 1994) as well as in Andromeda, and it seems to be missing as well in all the other spiral galaxies inside which black holes are assumed to exist. Therefore, besides the first unproved assumption of black holes as the source of anomalous gravitational effects, a second assumption for the conventional concept has to be accepted in order to explain these effects: that these black holes are surprisingly dormant.

Further strong arguments against black holes as the origin of the anomalous gravitational effects may be derived from the rotational velocity curves measured close to the centers of galaxies. From a black hole, a Keplerian dependence proportional to $r^{-.5}$ of the rotational velocity v on the distance r from the center is expected up to distances where the luminous matter contributes appreciably to the gravitational effects. On the basis of the alternative reflections on gravitation, however, a monotonous increase of the rotational velocity, starting from a value zero at the center, is expected due to the amount of the luminous matter and the gravitational factor $G_f$.

A very large amount of radius dependent rotational velocity data has been published (Persic, Salucci, & Stel 1996) almost all of which show a linear increase with distance from the center, especially for spiral galaxies. However, the angular resolution of these older rotational velocity data typically does not allow a decision whether a black hole may exist at the center or not. The above mentioned paper by Kormendy at al. (1998), however, contains not only the velocity dispersion data which are used to derive the existence of a black hole with a mass of $1.8 \times M_{Sun}$, but also presents rotation velocity data with sufficiently high resolution. Around 1 arcsec, the observed rotation curve shows a maximum of about 110 km/s. This value is rather well compatible with the orbital velocity around a black hole with the above given mass of $1.8 \times M_{Sun}$. The angular resolution of 0.2 arcsec allows the determination of the rotation velocity v at distances far below 1 arcsec. Below 1 arcsec, the experimental data do not show the Keplerian behavior expected from the existence of a black hole. Instead, a rather linear increase from the center with a slope slowly decreasing with radius r is observed. This is expected if, in equation (10), G is replaced by $G_f$ in the product of $M(r)G$. According to equation (7) with the exponent a = 1, the factor $M(r)$ increases proportional to r, but the logarithmic decrease of $G_f$ according to Fig. (4) reduces this slope with increasing r, in accordance with the observations. The authors concede that the non-Keplerian rotation velocity curve is a shortcoming which deserves further exploration, but they also note that this behavior is not impossible in principle.

According to Tsiklauri & Viollier (1998), „the identification of a central supermassive object in the Milky Way has been a source of continuous debate in the literature" and, as they state, there is no compelling proof that supermassive black holes actually do exist" in galactic centers. In order to explain the so called blackness problem which follows from the assumption of a massive black hole at the center especially of the own Galaxy, they propose a new explanation: a supermassive object composed of self-gravitating, degenerate heavy neutrinos. Yet this alternative to the black hole interpretation, which of course assumes a more extended object, still awaits a definite proof as well.

## 6. CONCLUSIONS

On the basis of previously published alternative reflections on gravitation, the existence of anomalous gravitational effects is postulated, especially within the regions of spherical systems with highly concentrated mass-accumulations such as globular star clusters, elliptical galaxies, bulges of spiral galaxies and clusters of galaxies. Indeed, the cited experimental observations show that such anomalous effects do exist within the regions of all four mentioned groups of mass accumulations. Especially within two groups, the bulges of spiral galaxies and the clusters of galaxies, where the gravitational effects are strong enough and where the relative limit of error is low, a very large number of objects is found to possess the postulated anomalous gravitational effects. In every case, the gravitational effects show the increase towards the center of the object and the rotational symmetry as expected in accordance with the considerations in chapter 3. Conventionally, these anomalous gravitational effects are explained on the basis of Newtonian mechanics by the assumption of the existence of black holes or dark matter with appropriate amounts and spatial distribution. Further assumptions, e.g. dormant states in the case of missing X-ray radiation from black holes, are taken into consideration if necessary. As long as these powerful concepts are the commonly accepted explanations of the anomalous gravitational effects, it is difficult to introduce new concepts like the alternative reflections on gravitation discussed here. In spite of this, as can be seen from this paper, this concept may have at least one useful application: to serve as a simple and very effective method to predict further locations in the universe where new members of the large list of exiting black holes may be found.

Modern powerful equipment such as the Hubble Space Telescope and the largest ground-based telescopes together with sophisticated experimental and computational techniques may in the near future deliver new cosmological results which allow a more definite decision between conventional Newtonian mechanics and the alternative reflections on gravitation. Possibly, one effect of this type has already been obtained prior to the completion of this paper: From the observation of distant supernovae, the unexpected conclusion has to be drawn that space itself seems to be permeated by a repulsive force and that cosmic expansion has not decreased but sped up during the last billions of years (Glanz 1998b). Just this effect, however, has to be expected on the basis of the alternative reflections (Albers 1997a) due to the primary, repulsive forces acting between all masses of the universe. Standard gravity theory, together with the big bang concept, did not predict such a result. The solution of this problem which has recently predominated the discussion in scientific journals, internet news and newspapers is the cosmological constant. Such a concept, introduced by Albert Einstein who himself called it his „greatest blunder", probably leads, together with the concepts of dark matter and black holes, to a survival of the standard gravity theory, at least in the near future.

**Acknowledgment:** The author gratefully acknowledges interesting and useful discussions with K. Zioutas about the dark matter problem during the preparation of the first paper (Albers 1997a) and the critical reading of all three manuscripts.

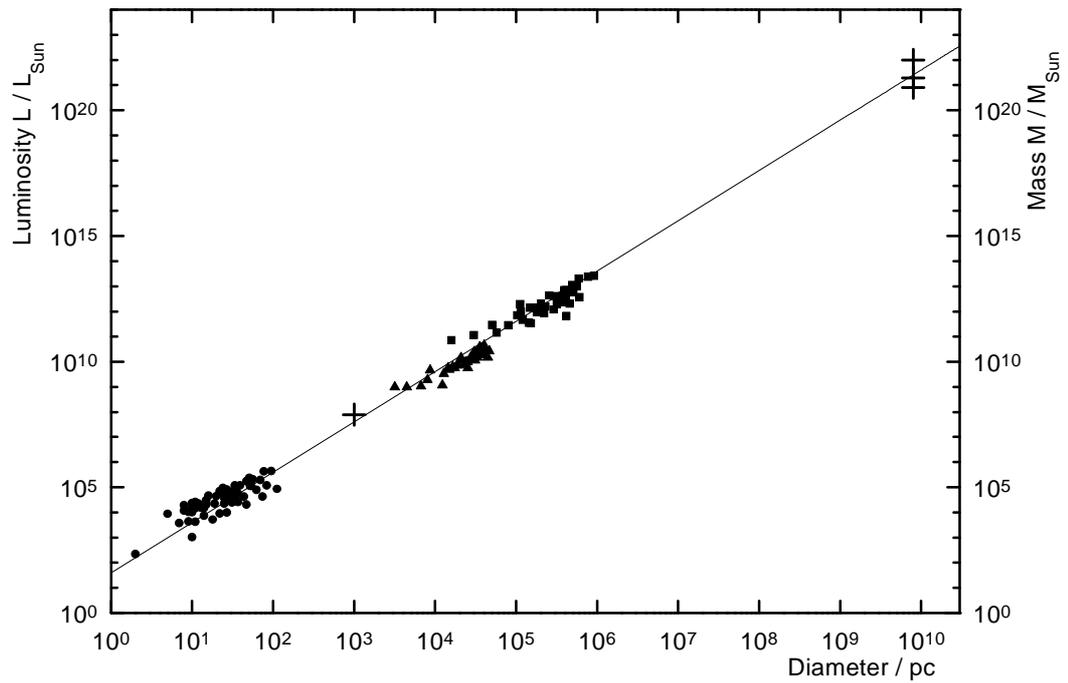

Fig. 1.- Dependence of luminosity L and mass M on the diameter of different spherical objects in the universe. Circles: globular star clusters; triangles: elliptical galaxies; squares: spherical clusters of galaxies (all data from Albers 1997b). The cross at 1000 pc represents the bulge of our Galaxy. The three crosses at a diameter of 8 Gpc correspond to the universe and result from different literature data of density or mass as described in the text.

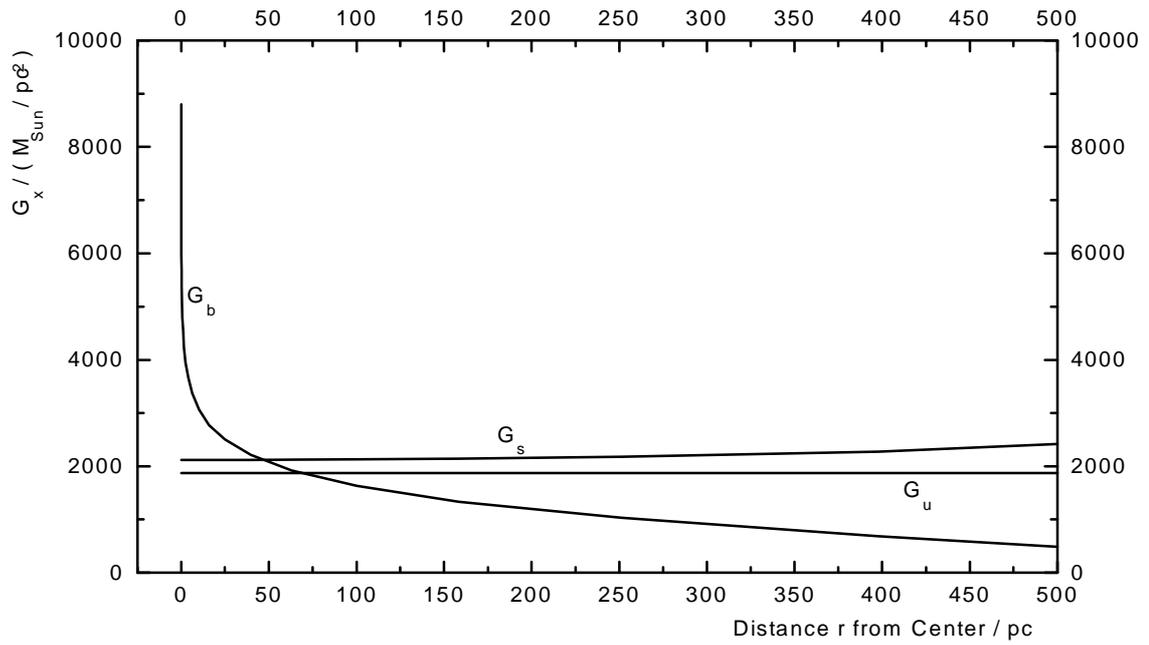

Fig. 2.- Increase of the components $G_x$ of the gravitational factor $G_f$ towards the center of the bulge with a radius $R_b$=500 pc inside a typical spherical galaxy. Different lines represent the bulge contribution, $G_b$, the contribution of the spiral system, $G_s$ and the contribution of the universe, $G_u$.

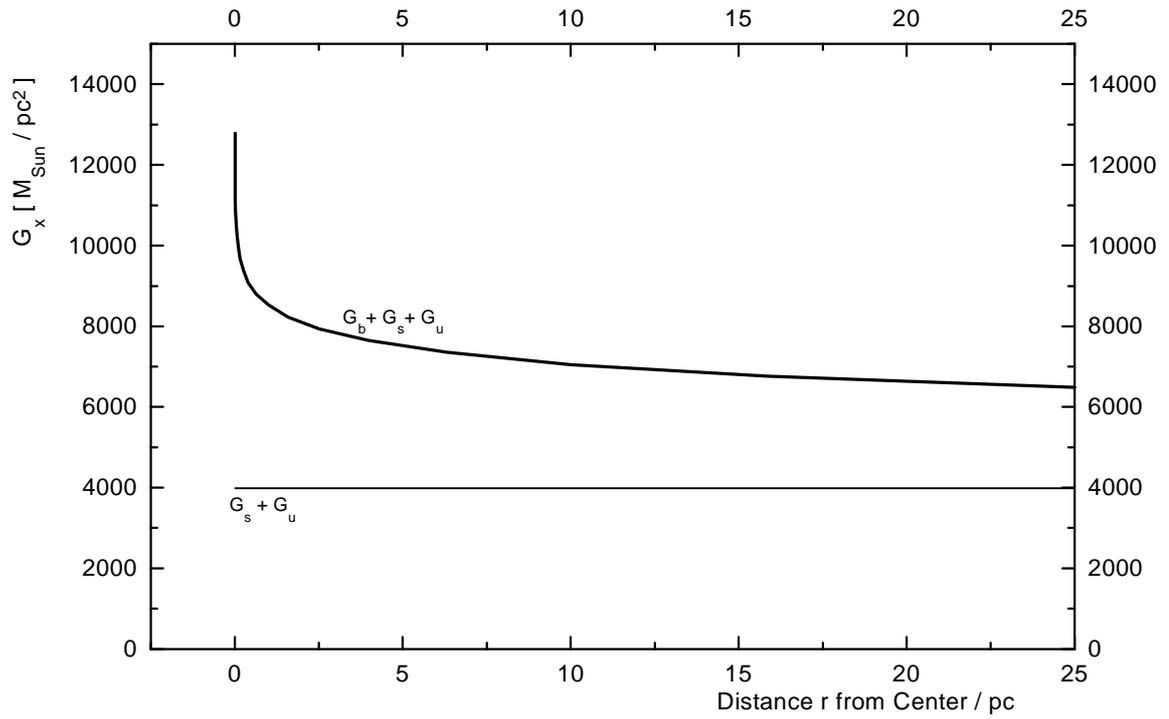

Fig. 3.- Dependence of the gravitational factor $G_f = G_b + G_s + G_u$ (contributions from the bulge, spiral system, and the universe) on the distance from the center of a bulge with radius R=500 pc in the innermost 25 pc.

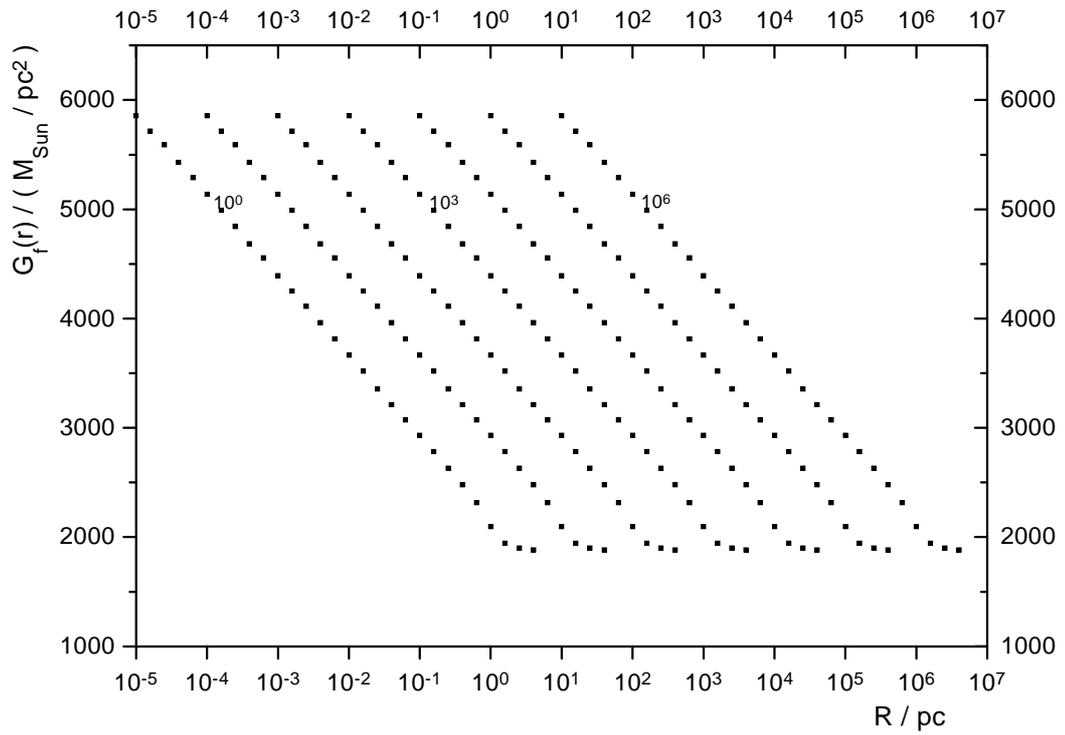

Fig. 4.- Dependence of the gravitational factor $G_f = G_{sp} + G_u$ (contributions from spherical systems and the universe) on the distance from the center of spherical systems with different radius R. R is drawn next to the individual curves and is given in pc.